\documentstyle{article}
\begin{document}
\LARGE
\begin{center}
\bf Quantum Creation of Topological Black Hole
\vspace*{0.6in}
\normalsize \large \rm

Zhong Chao Wu

Dept. of Physics, Beijing Normal University

Beijing 100875, China

and

Dept. of Applied Mathematics, University of Cape Town 

Rondebosch 7700, South Africa

\vspace*{0.4in}
\large
\bf
Abstract
\end{center}
\vspace*{.1in}
\rm
\normalsize
\vspace*{0.1in}

The constrained instanton method is used to study quantum creation of
a vacuum or charged topological black hole. At the $WKB$ level, the relative creation probability is the exponential of a quarter  sum of the horizon areas associated with the seed instanton. 

\vspace*{0.3in}

PACS number(s): 98.80.Hw, 98.80.Bp, 04.60.Kz, 04.70.Dy


\vspace*{0.3in}

e-mail: wu@axp3g9.icra.it

\pagebreak

\vspace*{0.3in}

The cosmological $C-$ metrics [1] contain  many topological black hole spacetimes. In this paper we shall study the quantum creation of a topological black hole. Since one can count the Schwarzschild black hole, for example, twice from two spatial infinities, one can interpret it as a pair of black. In the no-boundary universe [2], we shall use the constrained instanton method to evaluate the relative creation probability [3][4]. The constrained instanton can mediate the creation of a black hole.
The regular instanton can be considered as a special case thereof.

The Euclidean metric of a nonrotating topological black hole can be written in a simple form
\begin{equation}
ds^2 = \Delta d\tau^2 + \Delta^{-1} dr^2 + r^2 d \Omega^2_2,
\end{equation}
where
\begin{equation}
\Delta = p - \frac{2m}{r} + \frac{Q^2}{r^2} - \frac{\Lambda r^2}{3}.
\end{equation}
Here, $m$, $Q$, $\Lambda$ are the mass parameter, magnetic or electric charge and cosmological constant, respectively, and $p$ is the sign of the constant curvature of the two dimensional space $d\Omega^2_2$ 
\[
d \Omega^2_2 =  d \theta^2 + \sin^2 \theta d \phi^2 \;\;\;\;\;\; p = 1,
\]
\[
d \Omega^2_2 =  d \theta^2 + d \phi^2 \;\;\; \;\;\;\;\;\;\;\;\;\;\;\;  p = 0,
\]
\begin{equation}
d \Omega^2_2 =  d \theta^2 + \sinh^2 \theta d \phi^2 \;\;\; p = -1.
\end{equation}
They correspond to the 2-sphere of genus $g = 0$, the torus of genus $g = 1$ and the 2-hyperboloid, respectively. The open 2-hyperboloid can be
compactified into a homogeneous space with genus $g \ge 2$ by identifying  opposite sides of an regular polygon of $4g$ sides with area $4\pi (g -1)$ on the 2-hyperboloid. The gauge field is
\[
F = Q \sin \theta d\theta \wedge d \phi \;\;\; \;\;\; p = 1,
\]
\[
F = Q d\theta \wedge d \phi \;\;\;\;\;\;\;\;\;\;\;\;\; p = 0,
\]
\begin{equation}
 F = Q \mbox{sinh} \theta d\theta \wedge d \phi \;\;\; p = -1
\end{equation}
for the magnetically charged case, and
\begin{equation}
F = \frac{ i Q}{r^2} d\tau \wedge dr
\end{equation}
for the electrically charged case. We do not consider the dyonic case.

At the $WKB$ level, the Lorentzian spacetime is created from a constrained instanton as the seed. It is a compact section of a complex manifold with a stationary action under the constraints that in the created universe a 3-geometry and the matter field
on it are given. For our case, we shall use the complex form of solution (1)-(5) to construct a compact section by cutting and pasting. In general, it will lead to some conical singularities. The validity  of the $WKB$ approximation requires the solution to have a stationary action. The existence of the singularity implies that the
manifold is not of a stationary action under no constraint. However, we shall show that its action is stationary under the constraints.

All Lorentzian sections obtained through analytic continuations from the constrained instanton can be interpreted as being created quantum mechanically from the seed. Its relative creation probability is
\begin{equation}
P \approx \exp - I_r,
\end{equation}
where $I_r$ is the real part of  the Euclidean version of the action $I$. It can written as
\begin{equation}
I = - \frac{1}{16 \pi} \int_M (R - 2 \Lambda - F^2) - \frac{1}{8
\pi}
\int_{\partial M } K,
\end{equation}
where $R$ is the scalar curvature of the spacetime $M$, $K$
is the trace of the second form of the boundary $\partial M$, and
$-F^2 = -F_{\mu \nu} F^{\mu \nu}$ is the Lagrangian of the Maxwell field.

We now construct the constrained instanton. One can factorize $\Delta$ into
\begin{equation}
\Delta (r) = - \frac{\Lambda}{3r^2} (r - r_0)(r - r_1)(r - r_2)(r - r_3),
\end{equation}
where $r_i\;\; (i = 0, 1,2,3)$ are zeros in the ascending order of their real parts.
We assume that, for physics motivation, among these roots, at least two roots are real and positive. This condition can be met in a range of these parameters. The other pair can be real or a pair of complex conjugates. If a pair of the roots are equal, or  complex conjugates, then the order is not essential. All these zeros are identified as horizons in a general sense.

The roots $r_l$ satisfy the following relations:
\begin{equation}
\sum_i r_i = 0,
\end{equation}
\begin{equation}
\sum_{i>j} r_i r_j = - \frac{3p}{\Lambda},
\end{equation}
\begin{equation}
\sum_{i>j>k} r_ir_jr_k = - \frac{6m}{\Lambda},
\end{equation}
\begin{equation}
\prod_i r_i = - \frac{3 Q^2}{\Lambda}.
\end{equation}

The  surface gravity $\kappa_i$ of $r_i$ is 
\begin{equation}
\kappa_i = \frac{\Lambda}{6r_i^2}\prod_{j = 0,1,2,3, \;\; (j
\neq i)}(r_i - r_j).
\end{equation}

The complex constrained gravitational instanton is formed by
identifying two sections of constant values of time $\tau$
between two complex horizons $r_i$ and $r_j$ [3]. The pasting introduces 
two $f_l$-fold $(l =i, j)$ covers around the horizons. The $f_l$-fold  cover then turns 
the $(\tau - r)$ plane into a cone with a deficit angle 
$2\pi (1-f_l)$ there.  Both $f_i$ and $f_j$ can take
any pair of complex numbers with the condition
\begin{equation} 
f_i \beta_i - (-1)^{i +j} f_j \beta_j = 0, 
\end{equation} 
where $\beta_l = 2\pi \kappa^{-1}_l $.
If $f_i$ or $f_j$ is different from  $1$, then the conical singularity contributes to the action expressed by a degenerate form of the surface
term in (7).

The singularity contribution to the action is
\begin{equation}
I_{h} = -\sum_{l = i, j} \epsilon \pi r_l^2(1 - f_l).
\end{equation}

The volume contribution to the action is
\begin{equation}
I_{v} = - \frac{\epsilon \Lambda \Delta \tau}{6} (r_j^3 - r_i^3) \pm \frac{\epsilon Q^2 \Delta \tau}{2} (r^{-1}_i - r^{-1}_j),
\end{equation}
where $\Delta \tau = f_i \beta_i$ is the identification time period in the pasting, $+ (-)$ is for the magnetic (electric) case, and $\epsilon$ is $1$
for case $p =1, 0$ and is $g-1$ for case $p = -1$. Here the unit torus area is chosen to be $4\pi$, for convenience.

One can use the joint section $\tau = 0$ and $\tau = \Delta \tau/2$ i.e. $\tau = - \Delta \tau/2$ as the equator, and then make a series of analytical continuations to obtain the wave function for the Lorentzian universe created.
The action form (7) is suitable for the wave function for a given magnetic charge, the charge is evaluated by the surface integral of the gauge field over $S^2$ $(\theta - \phi)$ space divided by $\epsilon$. 

From Eqs. (14)(15)(16) one obtains the total action
\begin{equation}
I =  -\epsilon \pi (r_i^2 + r_j^2),
\end{equation}
which is independent of the parameter $\Delta \tau$. $\Delta \tau$ is the only degree left under the restriction  that the 3-metric $h_{ij}$ and the magnetic charge at the equator are given. Therefore, the manifold pasted is qualified as a constrained instanton.

However, the configuration of the wave function at the equator is not proper for the electrically charged case. Instead, the Maxwell part of the action in (7) is set for the variation under the condition that the following variable conjugate to the charge is fixed
\begin{equation}
\omega =\epsilon \int A,
\end{equation}
where the integral of the vector potential $A$ is around the $S^1$ direction of the equator in the $(\tau -r)$ space. Or equivalently, the wave function obtained from the path integral using the action (7) is $\Psi (\omega, h_{ij})$. The most
convenient choice of the gauge potential for the calculation is
\begin{equation}
A = -\frac{iQ}{r^2}\tau dr.
\end{equation}

In order to obtain the wave function $\Psi (Q, h_{ij})$ for a given electric charge, one has to appeal to a representation transformation
\begin{equation}
\Psi (Q, h_{ij}) = \frac{1}{2\pi} \int^{\infty}_{-\infty} d
\omega e^{i\omega Q} \Psi
(\omega, h_{ij}).
\end{equation}
This Fourier transformation is equivalent to a multiplication of
an extra factor
\begin{equation}
\exp \left (\frac{- \epsilon \Delta \tau Q^2  ( r_i^{-1} -
r_j^{-1})}{2}\right )
\end{equation}
to the wave function. Or equivalently, this introduces an extra term into the action, which turns the $-$ sign in the matter term of (16) to $+$, and the effective action takes the same form as that for the magnetic case. One of the motivations for the Fourier transformation is to recover the duality between magnetic and electric black hole creations [3][4][5][6].

We have closely followed [3][4] in  deriving (17), using (9)(12). It is noted that the derivation of the result is independent of the condition
(10), which is the only place $p$ is involved. It is not surprising that the action should be a linear function of $\Delta \tau$. What is surprising is that the action is independent of $\Delta \tau$. Consequently, the action is the negative of the entropy associated with the two horizons of the instanton [7]. Our approach confirms that the black hole entropy is one quarter their horizon area [8]. This approach is not sensitive to the procedure used in the background subtraction [9].

In the no-boundary universe, the true constrained instanton for the same universe created should have the largest action in comparison with the rest of the instantons [2][4]. 

Therefore, for the case $\Lambda > 0$, if $Q = 0$,  $r_1$ disappears and one has to use the instanton with horizons $r_2, r_3$, these horizons are identified as the black hole and cosmological horizons. If $Q \neq 0$, one has to use the inner and outer black hole horizons $r_1, r_2$ for the instanton.   

For the case $\Lambda < 0$, one has to use the instanton with complex conjugate horizons $r_0, r_1$. $r_2$ and $r_3$ are the inner and outer black hole horizons.  If $Q = 0$, then the inner horizon $r_2$ disappears.

From Eqs. (9)(10) one finds that the sum of all horizon areas is a constant
\begin{equation}
\sum_{i = 0,1,2,3} S_i = \sum_{i = 0,1,2,3} 4\pi\epsilon r^2_i = \frac{24\pi \epsilon p}{\Lambda}.
\end{equation}

Since the relative creation probability is, at the $WKB$ level, the exponential
of the negative of the action, or of the entropy. For the case 
$\Lambda > 0, Q = 0\;\; (Q \neq 0)$, it is the exponential of a quarter of the sum of the outer black hole and cosmological (inner black hole) horizon areas. For the case $\Lambda <0, Q =0 \;\;(Q \neq 0)$, one can use (22) and conclude that
the relative probability is the exponential of a negative quarter of the black hole horizon area (a quarter of the negative sum of inner and outer black hole horizon areas). These results are very similar to those of their cousins, the Kerr-Newman-(anti-)de Sitter black holes [3][4].

We are also interested in the creation of black hole with noncompact horizons. One can begin with metric (1) for $p = -1$. Instead of the previous compactification approach, one can make an analytic continuation letting $\chi = i \theta$, then the metric takes the form
\begin{equation}
ds^2 = \Delta d \tau^2 + \Delta^{-1}dr^2 - r^2(d \chi^2 + \sin^2 \chi d \phi^2),
\end{equation}
The constrained instanton is similar to that for case of $S^2$ horizon with the modified signature. The creation probability should be the same as that for the case of compactified 2-hyperboloid horizon with $g= 2$.    

At the $WKB$ level, one can obtain the total creation probability of the  black hole for $\Lambda < 0$ with given charge and mass and all different topologies
\begin{equation}
P = P_1 + P_0 + P_{-1}\left ( 1 + \frac{1}{1- P_{-1}} \right ),
\end{equation}
where $P_1, P_0$ and $P_{-1}$ are the probabilities for the cases of horizons with topology 2-sphere, 2-torus and compactified 2-hyperboloid of $g=2$, respectively. $P_{-1}$ is also the creation probability included for the case of open 2-hyperboloid horizon. The fraction term accounts for all cases of a compactified 2-hyperboloid horizon.

The case $\Lambda =0$ can be considered as the limiting case as $\Lambda$ approaches zero from below. 

All known results of topological black hole creations mediated by regular instantons [10] become special cases of the consideration here. It is noted that for the black hole creation in the open space background we do not require the presence of domain wall for the compactification of spacetime as in [10][11].

We have investigated the quantum creation of the black hole with 2-dimensional horizon of topology of genus $g$. There exists another kind of constant curvature black hole. This is  called the 4-dimensional version of the $BTZ$ black hole [12][13]. The case of the original 3-dimensional $BTZ$ black hole and its higher dimensional version is dealt with in a separate  publication.

\vspace*{0.4in}

\bf Acknowledgements
\rm
\vspace*{0.1in} 

I would like to thank G.F.R. Ellis of University of Cape Town for his hospitality.
\vspace*{0.1in}
  
\bf References:

\vspace*{0.1in} 
\rm

1.  J.F. Plebanski and M. Demianski, \it Ann. Phys. \rm \underline{98}, 98 (1976).

2. J.B. Hartle and S.W. Hawking, \it Phys. Rev. \rm \bf D\rm
\underline{28}, 2960 (1983).

3. Z.C. Wu, \it Int. J. Mod. Phys. \rm \bf D\rm\underline{6}, 199
(1997), gr-qc/9801020.

4. Z.C. Wu, \it Phys. Lett. \bf B\rm 
\underline{445}, 274 (1999); gr-qc/9810077.

5. S.W. Hawking and S.F. Ross,  \it Phys. Rev. \bf D\rm 
\underline{52}, 5865 (1995).

6. R.B. Mann and S.F. Ross, \it Phys. Rev. \bf D\rm
\underline{52}, 2254
(1995).  

7. Z.C. Wu, \it Gene. Relativ. Grav. \rm \underline{31}, 1097 (1999), gr-qc/9812051.

8. D. Brill, J. Louko and P. Peldan, \it Phys. Rev. \bf D\rm \underline{56}, 3600 (1997). 

9. L. Vanzo, \it Phys. Rev. \bf D\rm
\underline{56}, 6475 (1997).

10. R.B. Mann, \it Nucl. Phys. \bf B \rm \underline{516}, 357 (1998).

11. P.R. Caldwell, A. Chamblin and G.W. Gibbons, \it Phys. Rev. \bf D\rm
\underline{53}, 7103 (1996).  

12. M. Ba$\tilde{n}$\rm ados, C. Teitelboim and J. Zanelli, 
\it Phys. Rev. Lett. \bf \rm \underline{69}, 1849 (1992).

13. M. Ba$\tilde{n}$\rm ados, 
\it Phys. Rev. \bf D\rm \underline{57}, 1068 (1998).

\end{document}